# Enhanced Modulation Technique for Molecular Communication: OOMoSK

Md. Humaun Kabir, and Kyung Sup Kwak

*Abstract* - **Molecular communication in nanonetworks is an emerging communication paradigm where molecules are used as information carriers. Concentration Shift Keying (CSK) and Molecule Shift Keying (MoSK) are being studied extensively for the short and medium range molecular nanonetworks. It is observed that MoSK outperforms CSK. However, MoSK requires different types of molecules for encoding which render transmitter and receiver complexities. We propose a modulation scheme called On-Off MoSK (OOMoSK) in which, molecules are released for information bit 1 and no molecule is released for 0. The proposed scheme enjoys reduced number of the types of molecules for encoding. Numerical results show that the proposed scheme enhances channel capacity and Symbol Error Rate (SER).**

## I. INTRODUCTION

Nanomachines, built from individual molecule or arranged set of molecules that are nanometer and micrometer in size, are able to perform tasks such as sensing, computing, storage and actuation [1]. Nanonetwork is the interconnection of nanomachines anticipated to perform collaborative tasks such as health monitoring, drug delivery, regenerative medicine, environment monitoring, waste/population control, pattern and structure formation etc [2]. Small-scale devices built from biological materials called bio-nanomachines, posing a high degree of energy efficiency and biocompatibility, are capable of interacting with biological molecules and cells in nano to micrometer scale. Transmitter nanomachines release the information-encoded molecules in the environment, the molecules propagate to receiver nanomachines, and the receiver nanomachines biochemically react with the molecules to decode the information [2]. The propagation of messenger molecules from transmitter to receiver nanomachine is governed by Brownian motion and is affected by two parameters: drift velocity and the diffusion coefficient [3].

This work was supported by the National Research Foundation of Korea (NRF) grant funded by the Korean government (MEST) (No.2010-0018116). The authors are with Inha University, South Korea. Email: hakim2021@yahoo.com.

In this letter, we consider free diffusion of the messenger molecules where no additional force is required for the propagation of messenger molecules.

CSK is analogous to amplitude shift keying in which, information symbols are encoded based on the concentration levels of the messenger molecules. in CSK, the higher the modulation order, the more the number of molecules is required for encoding. No fixed threshold is there even though the same type of molecules is used. In MoSK, bearing a resemblance to frequency shift keying, different types of molecules are used for encoding [4]. For $M$-ary MoSK, $M$ different types of molecules are required and the threshold points are different for different types of molecules. Hence, the complexity of the transmitter and receiver nanomachines increases as the modulation order increases. Taking these problems under consideration, we propose a modulation technique which requires reduced number of the types of molecules for encoding.

## II. SYSTEM MODEL

We consider a time-slotted molecular communication system with slot duration $T_s$ where the transmitter and receiver nanomachines are in a stationary fluidic medium at a distance $r$ apart. We, further, consider that the transmitter and receiver nanomachines are perfectly synchronized and the receiver is able to distinguish different types of messenger molecules. Symbols are assumed to be transmitted upon On Off Keying (OOK) modulation from the transmitter nanomachine; an impulse of $n$ molecules are released at the start of the slot for information bit 1 and no molecules is released for 0. Messenger molecules, with a common diffusion coefficient D, diffuse through the medium and the channel is considered to be memoryless. $D$, measured in $m^2/s$, describes the tendency of propagation of the messenger molecules through the medium and can be written as

$$D = \frac{k_B T}{b} \quad (1)$$

where $k_B$ is the Boltzmann constant in $JK^{-1}$, $T$ is the absolute temperature of the environment in Kelvin, and $b$ is the drag constant of the messenger molecule inside the given fluid, which depends on the

characteristics of both the messenger molecule ($S_m$) and fluid molecule ($S_f$). This constant $b$ is

$$b = \begin{cases} 4\eta\zeta_s & if \; S_m \approx S_f \\ 6\eta\zeta_s & if \; S_m \gg S_f \end{cases} \quad (2)$$

where, $\eta$ is the viscosity of the fluid and $\zeta_s$ defines the Stokes' radius of the propagating molecule [5].

The receiver nanomachine, equipped with receptors, counts the total number of molecules received during the time slot to decode information under the following decision rule:

$$N \underset{0}{\overset{1}{\gtrless}} z \quad (3)$$

where $N$ is random number indicating received number of messenger molecules and $z$ is threshold number of molecules. It should be noted that the molecule is removed once it is received by the receiver. We assume that all the receptors are identical, observe the same concentration, and act independently.

## III. PROPOSED MODULATION TECHNIQUE: OOMoSK

In OOMoSK, different information bits in a symbol are encoded onto different types of molecules and the molecules are released as a mixture. Molecules are released if the information bit is 1 and no molecule is released for 0. Let us consider an M-ary OOMoSK, as shown in Fig. 1. There are $k = \log_2 M$ bits in a symbol and, therefore, $k$ different types of molecules are required for encoding a symbol. At first, information bits in a symbol are arranged from serial to parallel. First bit is encoded onto first type of molecules, second bit onto second type and so on. When the first bit arrives at the first branch, it checks whether the bit is 1 or 0. If the information bit is 1, molecules of type 1 are picked from a molecule bank and are sent to the accumulator. If the information bit is 0, no molecule is sent to the accumulator. Similarly, for the second bit, molecules of type 2 are taken and are sent to the accumulator if the information bit is 1 and no molecule is taken for 0. The process goes on upto $k$-th bit of the symbol. The molecules are, then, released by the transmitter nanomachine (emitter) after accumulating all the molecules from all $k$ bits. It should be noted that, as expected, the transmitter remains off if all the information bits in a symbol are 0. The receiver nanomachine is equipped with different types of receptors for different types of molecules released by the transmitter nanomachine. By reacting biochemically, receptor of type 1 detects molecule type 1, receptor of type 2 detects molecule type 2 and so on, as shown in Fig. 2. The decision is made by counting the total number of corresponding molecules received. $l$-th information bit is 1 if $N_l \geq z_l$ and 0 if $N_l < z_l$, where, $N_l$ is a random number representing the total number of the molecules received, $z_l$ is the threshold number of molecules, and $l = 1, 2, ..., k$. The information bits are, then, arranged from parallel to serial to obtain output data. Fig. 3 depicts 4-ary OOMoSK for an input sequence of 11100100.

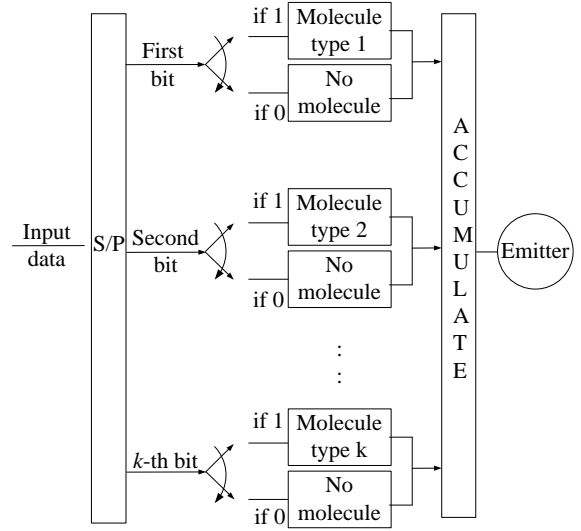

**Fig. 1:** Modulation

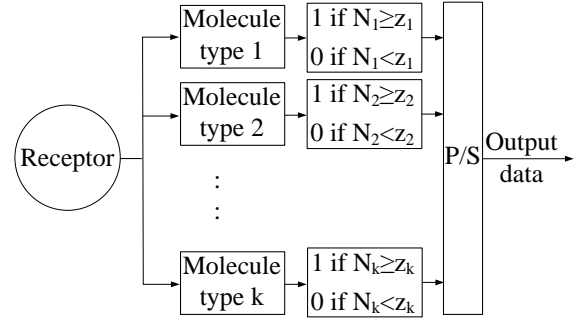

**Fig. 2:** Demodulation

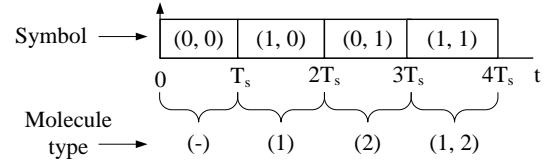

**Fig. 3:** 4-ary OOMoSK for an input sequence of 11100100.

## IV. PERFORMANCE ANALYSIS

The diffusion of molecules follows a probabilistic behavior. The probability density function of the time a messenger molecule requires to reach the receiver nanomachine called first hitting time, can be written as [6]

$$f(t) = \begin{cases} 0 & (t = 0) \\ \frac{r}{\sqrt{4\pi D t^3}} e^{-\frac{r^2}{4Dt}} & (t > 0) \end{cases} \quad (4)$$

By using (4), the probability that a molecule reaches the receiver nanomachine within the current slot $(\tau, \tau + T_s)$ is

$$p = \int_{\tau}^{\tau+T_s} f(t) dt \quad (5)$$

where, $\tau$ is a given transmission time. The number of received molecules $N$ during $(\tau, \tau + T_s)$ is a random variable and can be modeled as a binomial distribution

$$N \sim Binomial\ (n, p) \quad (6)$$

If the number of messenger molecules emitted by the transmitter is much larger (i.e $n \geq 1$), then the Binomial distribution of (6) can be approximated by Gaussian distribution $\mathcal{N}(\mu, \sigma^2)$ whose mean and variance are as follows:

$$\mu = np, \quad \sigma^2 = np(1-p). \quad (7)$$

### A. Types of Molecules

We notice from Fig. 1 that, for $M$-ary symbols, $k$ types of molecules are required for encoding a symbol in OOMoSK while in MoSK, in contrast, $2^k$ types of molecules are required.

### B. Energy Efficiency

Let us consider that, $n$ molecules are released for a $k$-bit symbol. Therefore, the number of molecules released for one bit is $\frac{n}{k}$. Total number of 1's in $M$ symbols in $M$-ary OOMoSK is $k2^{k-1}$. Therefore, total number of molecules released in OOMoSK for $M$ symbols is

$$= 2^{k-1} \times n \text{ molecules} \quad (8)$$

Similarly, for MoSK, total number of molecules released for $M$ symbols is

$$= 2^k \times n \text{ molecules} \quad (9)$$

From (8) and (9) we observe that, for $M$ symbols, the number of molecules released in OOMoSK is half the number of molecules released in MoSK. Therefore, if we assume the number of transmitting molecules for a symbol as the energy counterpart of electromagnetic communication then, the gain in energy in OOMoSK over MoSK is 50%.

### C. Symbol Error Rate

Let us consider $p_i = P(e|s_i)$ is the error when the symbol $s_i$ is sent. Therefore, the total symbol error probability

$$P_s = \sum_{i=0}^{M-1} q_i p_i, \quad M = 2^k \quad (10)$$

where $q_i$ is *a priori* probability of transmitting $i$-th symbol. Let us consider 4-ary OOMoSK, as shown in Fig. 3. One symbol is of two bits and two types of molecules are needed to transmit one symbol. We denote $s_0 = (0,0)$, $s_1 = (0,1)$, $s_2 = (1,0)$, and $s_3 = (1,1)$. For page limit, we show the calculation for $s_3$. $n_1 + n_2$ molecules are released at the start of the slot for the symbol $s_3$ where, the subscripts 1 and 2 in $n_1$ and $n_2$ represent the types of the molecules for first and second bits, respectively. We consider that two types of molecules are of two different sizes and $n_1 = n_2$. Since the molecules of different sizes will experience different diffusion coefficients, the threshold number of received molecules is different for different types of molecules. The number of received molecules $N_1$ and $N_2$ are given by $N_1 \sim Binomial\ (n_1, p_1)$ and $N_2 \sim Binomial\ (n_2, p_2)$, respectively, where $p_1$ and $p_2$ are the respective probabilities that a molecule of types 1 and 2 reaches the receiver nanomachine within the current slot $(\tau + 3T_s, \tau + 4T_s)$. We consider that the receiver can distinguish between the molecules of type 1 and 2. Therefore, the probability of error given $s_3$ is transmitted is

$$P(e|s_3) = P(s_0|s_3) + P(s_1|s_3) + P(s_2|s_3)$$
$$= P(N_1 < z_1).P(N_2 < z_2) + P(N_1 < z_1).P(N_2 \geq z_2) + P(N_1 \geq z_1).P(N_2 < z_2)$$
$$= P(N_1 < z_1) + P(N_1 \geq z_1).P(N_2 < z_2)$$

Because, $P(N_2 < z_2) + P(N_2 \geq z_2) = 1$. Therefore, the above equation becomes

$$P(e|s_3) = (1 - Q(U_1)) + Q(U_1).(1 - Q(U_2))$$
$$= 1 - Q(U_1).Q(U_2) \quad (11)$$

where $P(s_i|s_j)$ denotes the probability of receiving $s_i$ given $s_j$ is transmitted, $z_1$ and $z_2$ are the threshold numbers of received molecules for molecule type 1 and 2, respectively, $Q(.)$ is the tail probability of the Gaussian probability distribution function, $U_1 = \left(\frac{z_1 - n_1 p_1}{\sqrt{n_1 p_1 [1-p_1]}}\right)$ and $U_2 = \left(\frac{z_2 - n_2 p_2}{\sqrt{n_2 p_2 [1-p_2]}}\right)$. Similarly,

$$P(e|s_0) = 0 \quad (12)$$
$$P(e|s_1) = 1 - Q(U_2) \quad (13)$$
$$P(e|s_2) = 1 - Q(U_1) \quad (14)$$

Therefore, using (10), total symbol error probability is

$$P_s = q(3 - Q(U_1) - Q(U_2) - Q(U_1).Q(U_2)) \quad (15)$$

We assumed that $s_0$, $s_1$ etc. are equally probable i.e. $q_i = q$. If we consider the molecules of different types are of the same size (hydrofluorocarbon for example) then, $z_1 = z_2$ and $p_1 = p_2$. Writing $U_1 = U_2 = U$, we get

$$P_s = q(1 - Q(U))(3 + Q(U)) \quad (16)$$

### D. Capacity

Capacity can be calculated by using Shannon's formula which, defines the capacity as the maximum mutual information $I(X; Y)$ between the transmitted symbol $X$ and the received symbol $Y$ as

$$C = \max_z I(X; Y) \quad (17)$$

with $(X; Y) = \sum_{X \in \{s_i\}} \sum_{Y \in \{s_i\}} P(X; Y) \log_2 \frac{P(X;Y)}{P(X)P(Y)}$;

where, $i = 0, 1, ..., M-1$. $P(X;Y)$ can be calculated by manipulating equations (11) - (14). The marginal probabilities $P(X)$ and $P(Y)$ can be obtained by using $P(X) = \sum_{Y \in \{s_i\}} P(X;Y)$ and $P(Y) = \sum_{X \in \{s_i\}} P(X;Y)$, respectively.

## V. NUMERICAL RESULTS

SER and capacity for various quadrature modulation schemes are presented in Fig. 4 and Fig. 5, respectively. For numerical calculation, we considered different types of molecules of the same size irrespective of the modulation schemes and 125 molecules/bit are used on average. The size of the messenger molecules are assumed to be comparable to that of the fluid molecules. We considered slot duration $T_s = 20\ \mu s$, $\tau = 2\ \mu s$, and $r = 20\ \mu m$. The threshold number of molecules is assumed to be the same ($z = 20$) for 4-ary OOMoSK and QMoSK. In QCSK, however, the threshold varied since the number of transmitting molecules is different for different symbols. SER of $10^{-15}$ is achieved $D = 13 m^2/s$ in 4-ary OOMoSK while, SER of $10^{-4}$ and $10^{-2}$ are achieved at $D = 25 m^2/s$ in QMoSK and QCSK, respectively. Initially, QCSK outperforms 4-ary OOMoSK and MoSK. However, 4-ary OOMoSK shows better performance compared to other two as the diffusion coefficient increases. Capacity of 0.5 symbols/s is achieved $D = 4 m^2/s$ in 4-ary OOMoSK while, the same is achieved at $D = 12.55 m^2/s$ and $D = 23.5 m^2/s$ in QMoSK and QCSK, respectively. The threshold number of received molecules depends upon the receiver sensitivity and the lesser the number of threshold, the better the performance is. $\tau$ is defined as the time by which, among $n$ molecules, the first molecule reaches the receiver. In this study, we assumed $\tau = 2\ \mu s$ without loss of generality. However, this might not be the case in practical scenario.

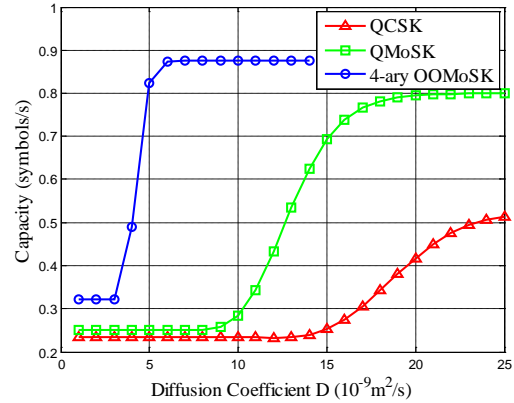

**Fig. 5:** Capacity comparison for different modulation schemes

## VI. CONCLUSIONS

Binary OOMoSK reduces to OOK and only one type of molecules is required for encoding. We have shown the results for 4-ary OOMoSK. However, the model can be extended to higher order. OOMoSK requires reduced number of the types of molecules for encoding which, helps reducing transmitter and receiver complexities. 4-ary OOMoSK outperforms QMoSK and QCSK. This is due to the fact that normalized signal energy, expressed as the average number of molecules per bit, is higher in OOMoSK than that of the other two schemes. We can infer that, for a target BER, our proposed scheme requires reduced number of molecules compared to other schemes.

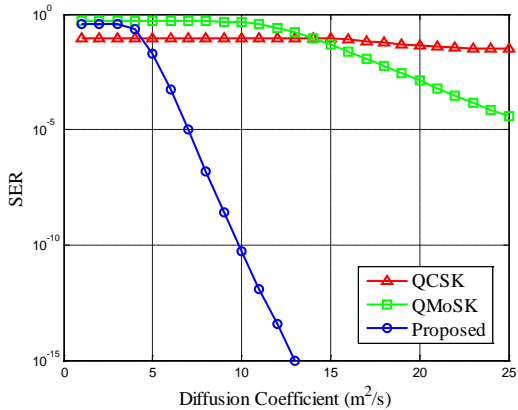

**Fig. 4:** SER comparison for different modulation schemes